\documentclass[conference]{IEEEtran}
\usepackage{cite}
\usepackage{amsmath,amssymb,amsfonts}
\usepackage{algorithmic}
\usepackage{graphicx}
\usepackage{subfigure}
\usepackage{textcomp}
\usepackage{xcolor}
\usepackage{makecell}
\usepackage{stfloats}
\usepackage{gensymb}
\usepackage{multirow}
\usepackage{caption}
\usepackage{algorithm}
\usepackage{algorithmic}

\usepackage{soul, color}

\def\BibTeX{{\rm B\kern-.05em{\sc i\kern-.025em b}\kern-.08em
    T\kern-.1667em\lower.7ex\hbox{E}\kern-.125emX}}
\begin{document}

\title{Emotion recognition based on multi-modal electrophysiology multi-head attention Contrastive Learning}

\author{
	\IEEEauthorblockN{1\textsuperscript{th} Yunfei Guo}\\
	\IEEEauthorblockA{\textit{Research institute} \\
		\textit{Chengdu Techman Software Co.,Ltd}\\
		Chengdu, China \\
		3180100017@caa.edu.cn}\\
	\and
	\IEEEauthorblockN{2\textsuperscript{th} Tao Zhang}\\
	\IEEEauthorblockA{\textit{Research institute} \\
		\textit{Chengdu Techman Software Co.,Ltd}\\
		Chengdu, China \\
		zhangtao@tme.com.cn}\\
	\and
	\IEEEauthorblockN{3\textsuperscript{th} Wu Huang{*}}\\
	\IEEEauthorblockA{\textit{School of Computer science} \\
		\textit{Sichuan University}\\
		Chengdu, China \\
		huangwu@scu.edu.cn}\\
}

\maketitle

\begin{abstract}
Emotion recognition is an important research direction in artificial intelligence, helping machines understand and adapt to human emotional states. Multimodal electrophysiological(ME) signals, such as EEG, GSR, respiration(Resp), and temperature(Temp), are effective biomarkers for reflecting changes in human emotions. However, using electrophysiological signals for emotion recognition faces challenges such as data scarcity, inconsistent labeling, and difficulty in cross-individual generalization. To address these issues, we propose ME-MHACL, a self-supervised contrastive learning-based multimodal emotion recognition method that can learn meaningful feature representations from unlabeled electrophysiological signals and use multi-head attention mechanisms for feature fusion to improve recognition performance. Our method includes two stages: first, we use the Meiosis method to group sample and augment unlabeled electrophysiological signals and design a self-supervised contrastive learning task; second, we apply the trained feature extractor to labeled electrophysiological signals and use multi-head attention mechanisms for feature fusion. We conducted experiments on two public datasets, DEAP and MAHNOB-HCI, and our method outperformed existing benchmark methods in emotion recognition tasks and had good cross-individual generalization ability. 
\end{abstract}


\section{Introduction}
Emotion recognition~\cite{kan2022self} refers to the use of computational techniques to identify and analyze human emotional states. This technology has applications in a variety of fields, including psychology, medicine, education, and social networking~\cite{koelstra2011deap}. In the medical field, emotion recognition can assist physicians in better understanding their patients’ emotional states, thereby improving treatment outcomes. Emotion recognition in education can assist teachers in assessing the emotional states of their students, enabling more impactful instruction. In social networking, emotion recognition can help platforms better understand their users’ emotional states, leading to improved service provision.

Electrophysiological signals, such as electroencephalogram (EEG), galvanic skin response (GSR)~\cite{healey2000wearable}, respiration rate (Respiration), and body temperature (Temperature), have certain advantages as input data for emotion recognition~\cite{huang2018multimodal}. These signals directly reflect physiological states and are closely related to emotional changes~\cite{healey2005detecting}. For instance, when an individual experiences tension or anxiety, their skin resistance decreases, their respiration rate increases, and their body temperature changes~\cite{zeng2007survey}. By measuring these signals, accurate inferences can be made about an individual’s emotional state.

However, there are also challenges associated with using electrophysiological signals for emotion recognition. Firstly, measuring these signals requires specialized equipment and expertise~\cite{soleymani2011multimodal}, which may increase costs and affect portability. Secondly, electrophysiological signals may be subject to interference or noise from external sources, necessitating preprocessing and filtering to improve signal quality~\cite{lazar2002combining}. Furthermore, the accuracy of emotion recognition may be influenced by physiological variations among individuals.

In summary, while the use of electrophysiological signals for emotion recognition has certain advantages, it also presents challenges. Future research must explore ways to overcome these challenges in order to improve the accuracy and reliability of emotion recognition.

Self-supervised contrastive learning~\cite{jaiswal2020survey} is an unsupervised learning method that learns the intrinsic structure of data by comparing the similarities and differences between different data samples. This approach can effectively leverage large amounts of unlabeled data to train deep neural networks, thereby improving the generalization and accuracy of the model. In emotion recognition, self-supervised contrastive learning can be used to learn emotion-related feature representations~\cite{taleb2022contig}, providing strong support for subsequent emotion classification~\cite{kan2022self}.

Multi-head attention~\cite{vaswani2017attention} is a mechanism for capturing long-range dependencies within sequential data. It computes attention weights between different positions in parallel using multiple “heads,” effectively capturing complex patterns in sequential data~\cite{tao2020eeg}. In emotion recognition, multi-head attention can be used to process time-series data such as speech signals, text data, or physiological signals to extract emotion-related features~\cite{sangeetha2021sentiment}.

In summary, self-supervised contrastive learning and multi-head attention have great potential for application in emotion recognition. Future research can further explore the use of these techniques in emotion recognition and combine them with other methods to improve the accuracy and reliability of emotion recognition.

The primary contributions and novelties of this paper encompass: 1) the application of self-supervised contrastive learning to cross-domain experiments with multimodal data fusion, exploring a novel approach to improve the generalization performance of cross-domain learning; 2) the effective extraction of task-based features from multimodal data through multi-head attention mechanisms, providing strong support for applications such as emotion recognition; and 3) the maintenance of good robustness on cross-data and cross-modal data with certain domain differences, illustrates the dependability and feasibility of the suggested approach in real-world scenarios. In summary, by applying self-supervised contrastive learning and multi-head attention mechanisms to multimodal data fusion and feature extraction, this paper provides new insights and methods for research in fields such as emotion recognition.

\section{Related work}
In previous research, emotion recognition using electrophysiology has been primarily divided into two categories: traditional feature engineering methods and deep learning approaches. Conventional feature engineering methods depend on manually designed features to represent electrophysiological signals. These features often comprise statistical measures (such as mean~\cite{koelstra2011deap}, variance~\cite{wang2006affective}, and skewness~\cite{soleymani2011multimodal}), frequency domain features~\cite{li2021mindlink} (e.g., power spectral density), and time-frequency features (e.g., wavelet transform~\cite{hazarika1997classification}~\cite{djafer2017identification}).
These methods often require domain knowledge and expertise to select appropriate features and may not fully exploit complex patterns in the data. Deep learning approaches automatically extract data representations through multi-layer neural networks~\cite{marjit2021eeg}. These methods do not require hand-crafted features and instead train models on large amounts of data to automatically discover relevant patterns. Deep learning methods have achieved significant success in emotion recognition, demonstrating excellent accuracy and robustness~\cite{zhang2022emotion}. In summary, emotion recognition methods based on electrophysiological signals include traditional feature engineering approaches and deep learning methods. Each approach has its strengths and weaknesses, and the choice of method depends on the specific application scenario and data characteristics.

In related work on electrophysiological data, researchers have often achieved performance improvements through the use of self-supervised contrastive learning algorithms, data augmentation, and targeted loss functions. Data augmentation is a technique used to expand the dataset by artificially generating new data samples through methods such as rotation, flipping, cropping, or adding noise~\cite{krishnan2022self}. In self-supervised contrastive learning, data augmentation can be used to generate positive and negative samples to help the model learn the intrinsic structure of the data. Meiosis~\cite{kleckner1996meiosis} is a genetics-inspired data augmentation method for the proposed Self-supervised Group Meiosis Contrastive learning~\cite{kan2022self}(SGMC) framework for emotion recognition. It leverages the alignment of stimuli between a set of EEG samples to generate augmented groups through pairing, crossover exchange and separation. The role of meiosis in self-supervised learning is to increase the meaningful difficulty for the model to decode EEG signal samples and mix signals from different subjects while preserving original stimulus-related features for SGMC extraction. Meiosis facilitates diversity exploitation of group composition through random pairing for crossover and separation. Overall, meiosis plays a crucial role in improving the performance of SGMC emotion recognition models, especially in label-scarce scenarios.

An important approach is to design loss functions specific to a given task, which measure the distance between predicted and actual values based on particular data augmentation methods. Common loss functions include mean squared error~\cite{sekar2020planning}, cross-entropy~\cite{zhang2021unleashing}, and contrastive loss~\cite{haochen2021provable}. In self-supervised contrastive learning, the loss function is used to guide model optimization to minimize prediction errors~\cite{misra2020self}.

In summary, in the analysis of temporal electrophysiological data, self-supervised contrastive learning provides new ideas and methods for applications such as emotion recognition by combining techniques such as meiosis data augmentation and loss functions.

Since ME signals are inherently long sequential data, it is theoretically feasible to extract cross-modal effective featur   es from ME data using multi-head attention mechanisms~\cite{song2018attend}. It computes attention weights between different positions in parallel through multiple "heads", effectively capturing complex patterns in sequence data. In ME data processing~\cite{wu2022video}, multi-head attention can be used to capture local dependencies between different modalities to extract task-relevant features.

In addition, multi-head attention has the advantages of enhanced expressiveness and improved computational efficiency. Since each "head" can learn different attention weights, multi-head attention can better express complex patterns in the data. At the same time, since multiple "heads" can be computed in parallel~\cite{kim2020multi}, multi-head attention can also improve computational efficiency.

In summary, in multimodal data processing, multi-head attention provides new ideas and methods for applications such as emotion recognition by capturing local dependencies, enhancing expressiveness and improving computational efficiency.

\section{Method}

\subsection{Overall Framework}\label{AA}

This paper implements emotion recognition using a multi-head attention self-supervised group meiosis contrastive learning framework for ME data based on domain differences.

\begin{figure*}[t]
\centerline{\rotatebox{90}{\includegraphics[scale=0.35]{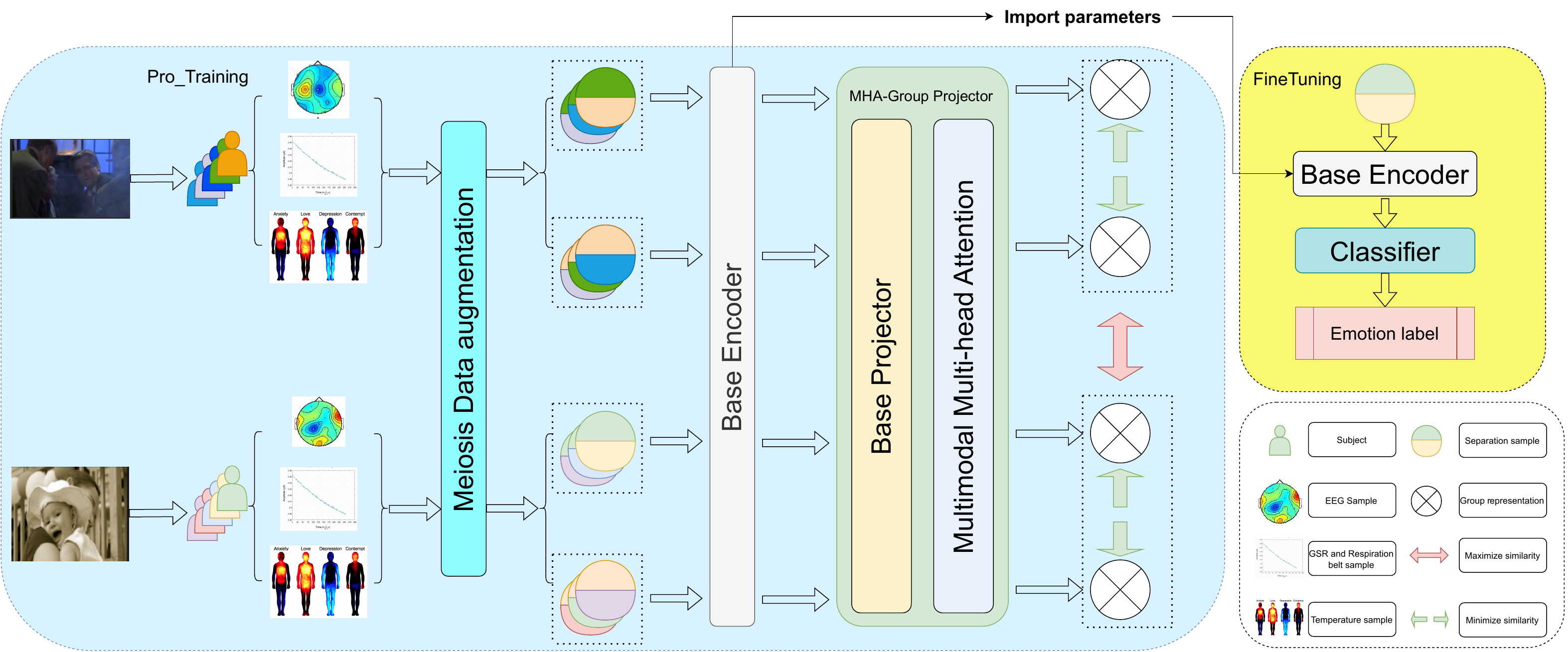}}}
\caption{The framework of ME\_MHAC algorithm. Fig. 1 shows in the model pre-training stage, ME signals corresponding to a video clip stimulus were sampled in groups. Then, the ME data of each group were processed by genetic method - meiosis, so as to generate positive and negative contrast data to achieve data feature enhancement. The further extended group uses the basic encoder to enhance the individual representation of individual samples of the subject, and then realizes the aggregate individual representation by the group projector to obtain group-level representation. This model needs to maximize the similarity between groups of the same video stimulus while minimizing group-level representations corresponding to different video stimuli.}
\label{fig}
\end{figure*}

 As shown in Fig. 1, the proposed framework consists of a contrastive learning pre-training stage and a model fine-tuning stage. The pre-training stage includes: a ME group sampler, meiosis data augmentation, a base encoder, a multi-head attention group projector, and a contrastive loss function. First, the ME group sampler generates mini-batch data by sampling from the ME data of the samples; secondly, meiosis is used to split and splice the ME signals of each group to generate two groups of ME signals to construct positive and negative ME signal pairs; thirdly, the base encoder extracts sample-level stimulus-related representations from each ME signal; then, the multi-head attention group projector aggregates the multimodal representations of each group to extract group-level video stimulus cross-modal related representations and maps them to the latent space; finally, the representations mapped to the latent space are optimized through the contrastive loss function for the parameters of the base encoder and group projector to achieve the purpose of minimizing contrastive loss. In the model fine-tuning stage, emotion recognition inference is performed using a pre-trained base encoder and an initialized classifier.

\subsection{ME Group Sampler}\label{BB}

Extracting features related to video stimuli from ME data and using contrastive learning algorithms for experimentation is challenging in terms of achieving data alignment. Therefore, this paper proposes the use of a ME data group sampler to obtain small batch data~\cite{koelstra2011deap}, providing a good foundation for subsequent data representation learning.

For the processed data, the video sequence number and subject are used as two tensor dimensions of the ME data, where each ME sample is defined as $ME^{s}_{v}\in R^{M*C}$, corresponding to the t-second ME signal recorded when subject s watches a t-second video clip v, where M represents the number of samples and C represents the number of channels used for ME data. To obtain a small batch of data, as shown in Fig.~\ref{fig:2}, the ME data sampler first randomly samples P video segments $v_1$, $v_2$,..., i.e. $v_P$ that has not been sampled in the current epoch. In order to extract two equal sample groups and construct positive pairs for each clip stimulus, the sampler then randomly selects 2Q subjects $s_{1}, s_{2},..., s_{2Q}$ for grouping. Further, the sampler extracts the ME data corresponding to the selected subjects and video segments, 2PQ samples $D = \{ME^{sk}_{vi} |i = 1,2,...P;K = 1,2,..., 2Q\}$, recorded respectively by 2Q subjects watching P video segments. In addition, we note that a group of samples $G_{i} = \{ME^{s_{1}}_{v_{i}}, ME^{s_{2}}_{v_{i}}\}$,..., $ME^{s_{2Q}}_{v_{i}}\}$ corresponds to video clip vi. In $G_{i}$, each individual sample shares similar related features. Thus, the sampler will provide P groups of samples ${G_{1}, G_{2},..., G_{P}}$ corresponding to P different pre-training stimuli.

\begin{figure}[htbp]
	\centering{\includegraphics[scale=0.66]{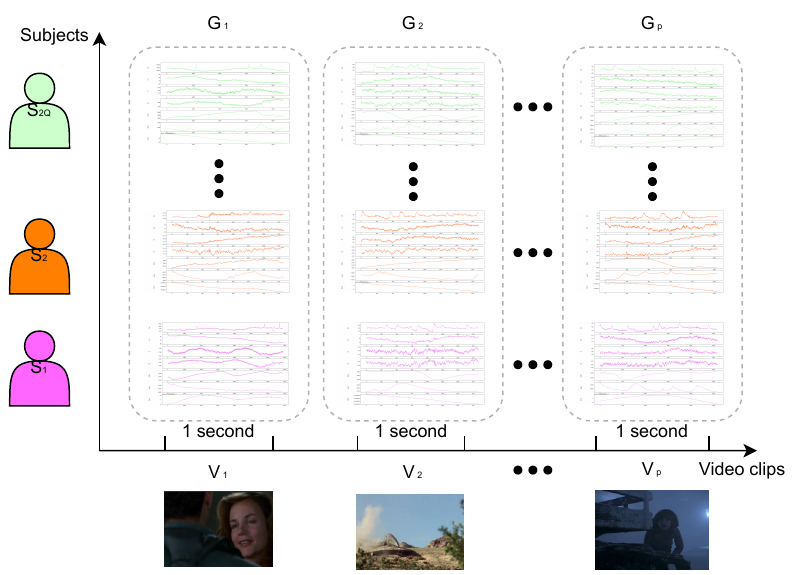}}
	\caption{Illustration of small batch sampling. The sampler first samples P video clips and 2Q subjects. For each sampled video clip, the sampler samples a set of ME signals recorded by the 2Q subjects as they watch. Then the ME sample samples of group P were obtained in small batches.}
	\label{fig}
\end{figure}

\subsection{Meiosis Data Augmentation}\label{CC}

Meiosis aims to build positive and negative sample pairs by utilizing the alignment of stimuli in ME groups, expanding a set of samples into two groups to maintain the same stimulus-related characteristics~\cite{koelstra2011deap}.

In order to increase the difficulty of the model in decoding the meaning of ME signal samples, we hope to mix signals from different subjects. In addition, in order to retain the original stimulus-related features extracted by ME-MHACL, we choose to split and splice signals corresponding to the same stimulus. Therefore, we designed the crossover transformation as follows: using $\{a_1, a_2,..., a_M\}$ to represent the ME signal A of any sample, where $a_i$ is the data of the i-th sampling point (i=1,2,...,M). Similarly, $\{b_1, b_2,..., b_M\}$ Further, we exchange the data of the first c sampling points of samples A and B to obtain $\tilde{A} = \{b_1, b_2,..., b_c, a_c+1, a_c+2,..., a_M\}$ and $\tilde{B}= \{a_1, a_2,..., a_c, b_c+1, b_c+2,..., b_M\}$, where c is known.This transformation of any two ME signals is encapsulated in the following function expression:

\begin{equation}
\{ \tilde A,\tilde B\}  = T\left( {A,B,c} \right)
\label{eq:1}
\end{equation}

In addition, to take full advantage of the diversity of group combinations, we can randomly pair for crossover and separation. As shown in Fig. 3, the overall design of meiosis data enhancement is as follows:

\begin{itemize}
    \item Individual pairing: For one original  ME signals group $G_i=\{ME^{s_1}_{v_i}|k = 1, 2, ..., 2Q\}$ (corresponding to a video clip $v_i$) individual ME signals are randomly paired to form $Q$ pairs $\{ME^{s_1}_{v_i}, ME^{s_1+Q}_{v_i}\}, \{ME^{s_2}_{v_i}, ME^{s_2+Q}_{v_i}\}$, ..., $\{ME^{s_Q}_{v_i}, ME^{s_2Q}_{v_i}\}$ for crossover.
    \item Crossover: Meiosis receives a randomly given split position $c$ to perform transformation (1) for each pairs to obtain $\{\{\tilde{ME}_{s_k}^{v_i}, \tilde{ME}_{s_k+Q}^{v_i}\}|k = 1, 2, ..., Q\}$.
    \item Separation: The transformed signals are randomly divided into two groups, and paired transformed signals are required enter into the different groups A and B. Two homologous groups of ME signals $\tilde{G}^{A}_{i} = \{\tilde{ME}^{s_k}_{v_i}|k = 1, 2, ..., Q\}$ and $\tilde{G}^{B}_{i} = \{\tilde{ME}^{s_k}_{v_i}|k = Q + 1, Q + 2, ..., 2Q\}$ can be obtained that sharing the similar group-level stimuli-related features. For the data expansion of ME data grouping samples, we use the following function expression:
\end{itemize}

\begin{equation}
\{ {\mathop {\tilde G}\nolimits_i^A ,\mathop {\tilde G}\nolimits_i^B } \} = Meiosis(\mathop {\tilde G}\nolimits_i^{} )
\label{eq:2}
\end{equation}

\begin{figure}[htbp]
\centerline{{\includegraphics[scale=0.21]{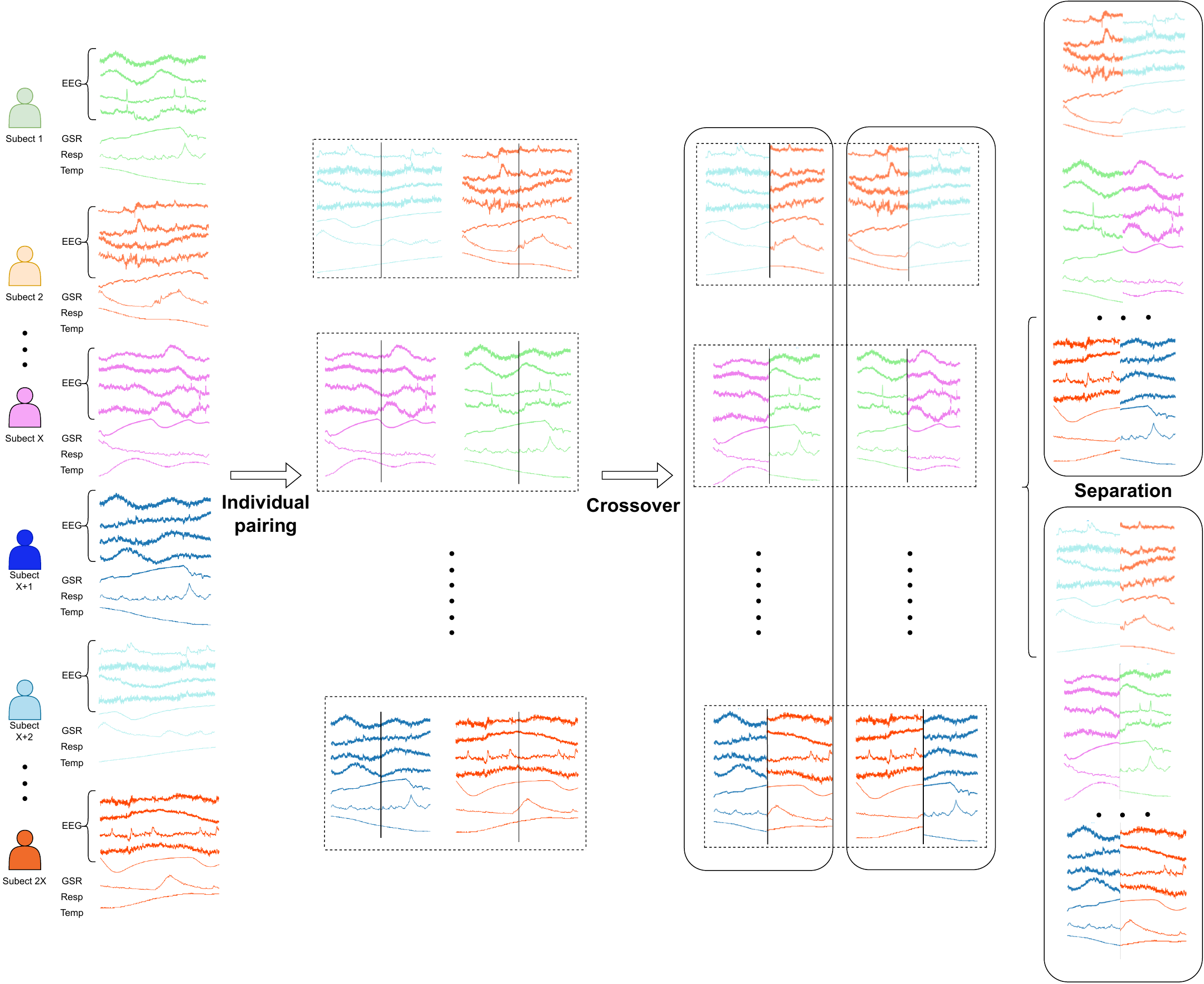}}}
\caption{Illustration of meiotic data enhancement. A group of ME samples sharing the same stimulus is randomly paired, part of the signal of one group is cross-exchanged, and then divided into two groups.}
\label{fig:3}
\end{figure}

When meiosis is established, for a minibatch of P group sample $\varsigma$, 2P group sample $\tilde \varsigma$can be obtained as follows:

\begin{equation}
\tilde \varsigma  = \left\{ {\tilde G_i^t|i = 1,2,...,P;t \in \left\{ {A,B} \right\}} \right\} = Meiosis(\varsigma )\
\label{eq:3}
\end{equation}

$\tilde{G}^A_i$ could from a positive pair with $\tilde{G}^B_i$, from negative pairs with any other 2(P-1) group samples.

\subsection{Base Encoder}\label{DD}

In order to extract group-level stimulus-related features for contrastive learning, a basic encoder was first designed to extract individual-level stimulus-related features from each individual ME sample. This paper introduces the basic encoder $f: R^{M \times C} \rightarrow R^D$, which maps individual ME samples $X$ to representations $h$ in a 512-dimensional feature space. Based on the existing model ResNet18-1D~\cite{cheah2021optimizing}, the basic encoder is designed as follows:

As shown in Fig. 4, it mainly consists of 17 Conv layers with 1D kernels. The first Conv layer has a kernel parallel to the time axis of the ME signal tensor, with a length of 9. Each residual block contains two Conv layers with the same number of kernels and length. In each residual block, the first layer’s kernel is parallel to the input ME tensor’s time axis, while the second layer’s kernel is parallel to the channel axis. For the 8 residual blocks, the kernel lengths decrease from large to small in the order of 15, 15, 11, 11, 7, 3, 3, and 5. The positions of the max pooling (Maxpool) with a 1D kernel, average pooling (Avgpool) with a 1D kernel, batch normalization (BN), and rectified linear unit (RELU) layers are shown in the figure.

\begin{figure}[htbp]
\centerline{{\includegraphics[scale=0.425]{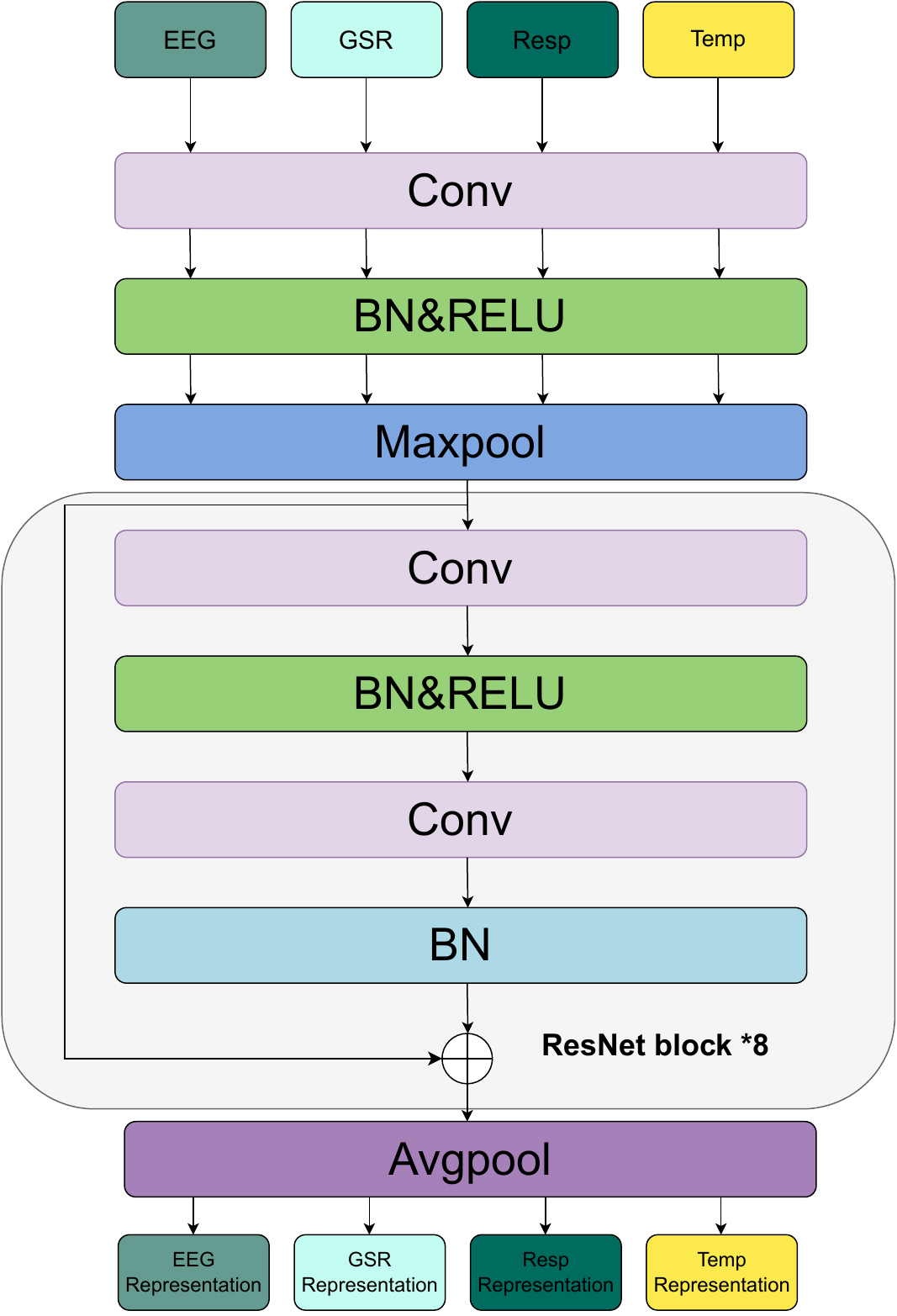}}}
\caption{Architectural details of the basic encoder. Conv represents the convolution layer by one-dimensional kernel. Maxpool and Avgpool reconfigure maximum pooling and one-dimensional core Avg pooling. BN stands for batch normalization. FC Layer indicates the fully connected layer. RELU stands for rectifiers linear unit.}
\label{fig:4}
\end{figure}

Through the basic encoder, for the augmented group sample $\tilde{G}^t_i$, the set of individual-level stimulus-related representations $\{h_1, h_2, \dots, h_Q\}$ can be obtained as follows:

\begin{equation}
H_{i_{ME}}^t = f(\tilde G_{i_{EEG}}^t,G_{i_{GSR}}^t,G_{i_{Resp}}^t,G_{i_{Temp}}^t)\
\label{eq:4}
\end{equation}

This collection is used to further extract group-level features. Individual representation can also be used to extract emotional features and classify emotions.

\subsection{Multi-head attention group projector}\label{EE}

The multi-head attention group projector aims to accurately project stimulus-related representations from ME signals into a latent space to compute the similarity of video clip stimuli. To alleviate the obstacles (fatigue, attention distraction, etc.) when extracting stimulus-related features from individual samples, a group projector was designed to extract group-level features from multiple samples.

A single ME sample set is a disordered matrix set, which lacks a specific extraction method. Most models focus on regular input representation, such as multi-channel images with a fixed order between different channels, and videos with a fixed order between different frames. \begin{figure}[htbp]
\centerline{{\includegraphics[scale=0.325]{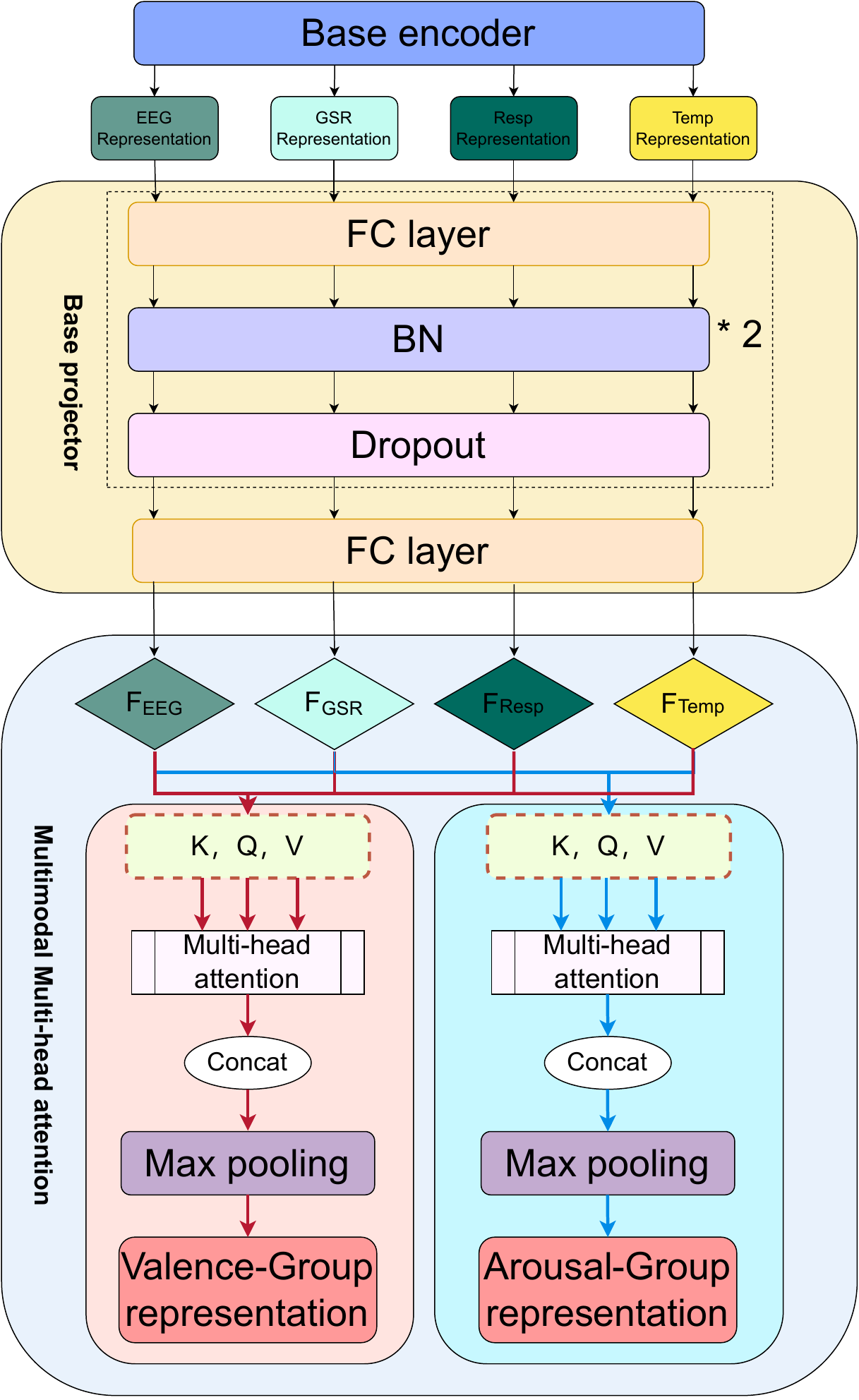}}}
\caption{Architecture details of the ME projection unit are as follows. FC layer represents a fully connected layer, while BN stands for batch normalization. The multi-head attention layer is placed after the third FC layer in the base projector. Maxpool is reset with maximum pooling. The binary classification representations based on Valence and Arousal are trained independently.}
\label{fig:5}
\end{figure}In the unordered point cloud classification problem, Charles R. Qi~\cite{qi2017pointnet} proposed PointNet, which uses a symmetric function to construct the network, achieving feature extraction of unordered point clouds.

To reduce the loss of individual features, extraction can be performed by increasing the dimension of individual representations. This paper proposes a basic projector $l: R^D \rightarrow R^H$, which uses a multi-layer perceptron (MLP) to project each individual representation $H$ onto a 4096-dimensional feature space. The basic projector consists of 3 fully connected layers, with hidden units decreasing from high to low in the order of 1024, 2048, and 4096. The activation functions for the first two layers use ReLU. The corresponding positions in the figure are Batch Normalization and Dropout set to 0.5.

The multi-head attention mechanism is a technique that can capture the local dependencies and global semantic information of input data. It divides the input data into multiple subspaces, calculates attention weights in each subspace, and then concatenates and linearly transforms the outputs of different subspaces to obtain the final output~\cite{ho2020multimodal}. The formula for the multi-head attention mechanism is as follows:

\begin{equation}
\text{F}(Q, K, V)_{i_{ME}}^t = \text{Concat}(head_1,\dots,head_h)H_{i_{ME}}^t\\
\label{eq:5}
\end{equation}

where each head is computed as:

\begin{equation}
\text {head}_i = \text{Attention}(QW_{i_{ME}}^Q, KW_{i_{ME}}^K, VW_{i_{ME}}^V)
\label{eq:6}
\end{equation}

and Attention is a scaled dot-product attention function:

\begin{equation}
\text{Attention} (Q,K,V) = \text{softmax}(\frac{{Q{K^T}}}{{\sqrt {{d_{head}}} }})V
\label{eq:7}
\end{equation}

Where $Q$, $K$, and $V$ represent the query, key, and value matrices, respectively, $d_{head}$ represents the dimension of each subspace, $head$ represents the number of subspaces, $W^Q_{i_{ME}}$, $W^K_{i_{ME}}$, and $W^V_{i_{ME}}$ represent the multimodal learnable representations obtained by transforming the group-level features $H^t_{i_{ME}}$ acquired through the basic encoder.

We employ an 8-head multi-head attention layer for feature fusion to obtain a comprehensive feature representation for emotion prediction. The multi-head attention mechanism takes the ME data as queries, keys, and values, and concatenates the output. In the binary classification tasks based on Arousal and Valence, as well as the quad-classification task introduced in the experiment, the attention of each modality is distributed differently due to task differences. Therefore, the model needs to be trained separately based on different labels to obtain the attention weights of different channels for each modality.

To ensure a constant output to represent a group sample with any input permutation, one-dimensional maximum pooling (MaxPool1D) is used to aggregate information from each dimension's upgrade representation. As shown in Fig. 5, MaxPool1D's 1D kernel is perpendicular to the dimensionally upgraded representation vector. The scan direction of the kernel is parallel to the upgraded representation vector, the step size is 1, and the fill is 0. The MaxPool can extract the maximum value of 4096 feature dimensions from Q-dimensional upgrade representation to obtain group-level feature representation in latent space.

We denote the group projection as $R^{Q \times D} \rightarrow R^H$. The group representation extracted in the latent space can be obtained through $g$:

\begin{equation}
\begin{split}
me_v^t &= g(\text{F}_{i_{ME}}^t)\\
       & = MaxPool1D(l({F_{1_{ME}}}),l({F_{2_{ME}}}), \ldots ,l({F_{Q_{ME}}}))\
\label{eq:8}
\end{split}
\end{equation}

Inspired by the above idea, we designed a model suitable for feature extraction of group ME signals using a symmetric function. As shown in Fig. 5, we designed a multi-head attention group projector composed of a basic projector, a multi-head attention module, and a symmetric function MaxPool1D.

\subsection{Classifier}\label{EE}

In the fine-tuning task of emotion classification based on arousal and valence, we use a classifier to extract emotional features from the representations extracted by the basic encoder and predict emotional labels. As shown in Table 1, the classifier mainly consists of three fully connected layers, with hidden units decreasing from high to low in the order of 512, 256, and 128. The corresponding positions in the figure are ReLU and Dropout set to 0.5.

\begin{table}[htbp]
\caption{Parameters of Classifier}
\begin{center}
\begin{tabular}{|c|c|c|c|}
\hline
\textbf{Model}&\multicolumn{3}{|c|}{\textbf{Parameters of layers}} \\
\cline{2-4} 
\textbf{structure} & \textbf{\textit{Layers}}& \textbf{\textit{Input features}}& \textbf{\textit{Output features}} \\
\hline
\multirow{2}{*}{Layer 1} & fc-classifier1& 512& 256 \\
\cline{2-4}
& bn-fc1& 256& 256 \\
\hline
\multirow{2}{*}{Layer 2} & fc-classifier2& 256& 128 \\
\cline{2-4}
& bn-fc2& 128& 128 \\
\hline
Layer 3& fc-classifier3& 128& 4 \\
\hline
\end{tabular}
\label{tab：1}
\end{center}
\end{table}

\subsection{Contrastive Loss}\label{GG}

To measure the similarity of group-level stimulus-related features between two groups of samples, we can calculate the cosine similarity of their group representation vectors. Given the input group samples $\{\tilde{G}^t_i |i = 1,2,\dots, P;t\in\{A, B\}\}$, we obtain the group feature representations $\{\tilde{me}^t_i |i = 1,2,\dots, P;t\in\{A, B\}\}$ through the basic encoder and multi-head attention group projector. Then, we can calculate the similarity between two augmented group samples $\tilde{G}^A_i$ and $\tilde{G}^B_j$ on $\tilde{me}^A_i$ and $\tilde{me}^B_j$:

\begin{equation}
\begin{split}
\text{sim} &=s(me_i^A,me_j^B) \\
		   &= \frac{{me_i^A \cdot me_j^B}}{{\left\| {me_i^A} \right\|\left\| {me_j^B} \right\|}},s(me_i^A,me_j^B) \in [0,1]\
\label{eq:9}
\end{split}
\end{equation}

The contrastive loss aims to maximize the similarity of group-level representations of two groups sharing the same stimulus label in a positive pair. 

\begin{equation}
\gamma  = {\Re _{[j \ne i]}} \in \left\{ {0,1} \right\}\
\label{eq:10}
\end{equation}

where $\gamma$ is an indicator function equaling to 1 if $j \ne i$. Similar to the SimCLR framework~\cite{pmlr-v119-chen20j}, we use normalized temperature-scaled cross-entropy to define the loss function as follows:

\begin{equation}
\ell _i^A =  - \log \frac{{\exp (\text{sim}/\tau )}}{{\sum\limits_{j = 1}^P {\gamma \exp (\text{sim}/\tau ) + \sum\limits_{j = 1}^P {\exp (\text{sim}/\tau )} } }}\
\label{eq:11}
\end{equation}

$\tau$ is the temperature parameter of softmax. The smaller the loss function is, the larger similarity between $me_i^A$ and $me_j^B$, and the similarity between $me_i^A$ and othor group representations come from the same minibatch.

Finally, the total loss for an iteration is the average of all contrastive losses for backpropagation as follows:

\begin{equation}
L = \frac{1}{{2P}}\left( {\ell _i^A + \ell _i^B} \right)\
\label{eq:12}
\end{equation}

\subsection{Pre-training Process}\label{HH}

ME-MHACL pre-training can be performed based on the constructed ME group sampler, meiosis data enhancement, basic encoder, multi-head attention group projector and contrast loss function.

In pre-training, we first set a number of epochs $T_1$, and then iterate over the epochs. In each iteration, we continue to sample $P$ video clips until all video clips are enumerated. In each iteration, the sampler extracts $2PQ$ ME samples $D = \{ME^{s_{k}}_{v_{i}} |i = 1,2,\dots,P;k = 1,2,\dots,2Q\}$ and combines them into groups $\varsigma = \{G_i |i = 1,2,\dots,P\}$.

Then, in the Meiosis data augmentation stage, to avoid deceiving the model by recognizing the splitting position, a fixed splitting position $c$ is randomly generated and sent to each Meiosis data of this iteration ($1 < c < M-1$). The $2Q$ augmented group samples $\tilde{\varsigma} = \{\tilde{G}_i^t | i = 1,2,\dots,P; t \in \{A,B\}\}$ can be obtained through equation (3). Further, we extract group-level features by fusing multimodal representations through the multi-head attention mechanism and projecting them into the latent space, obtaining group representations through equations (4)-(8). Further, we calculate the loss $L$ through equations (9)-(12). Finally, we minimize the loss $L$ through back propagation to compute gradients for updating the parameters of $f$ and $g$ using an optimizer. The specific steps are summarized in Algorithm 1.

\begin{algorithm}
\caption{Multi-modal Electrophysiology Multi-head Attention Contrast Learning}
\label{alg:example}
\begin{algorithmic}[1]
\REQUIRE Number of video clips $P$ per minibatch, number of subjects $Q$ per group. Initialized base encoder $f$ and group projector $g$. $ME$ represents ME data.
\FOR{epoch = 1 to $T_1$}
    \REPEAT
        \STATE Sample $P$ video clips $\{v_i|i = 1, 2, ..., P\}$.
        \STATE Randomly select $2Q$ subjects $\{s_k|k = 1, 2, ..., 2Q\}$.
        \STATE Sampler pack minibatch $G = \{G_i|i = 1, 2, ..., P\}$ from $D = \{ME^{s_k}_{v_i}|i = 1, 2, ..., P; k = 1, 2, ..., 2Q\}$
        \STATE Randomly generate a split position $c$.
        \STATE Obtain $\tilde{\varsigma} = \{ {\tilde{G}^{t}_{i}}|i = 1, 2, ..., P;t \in \{A, B\}\}$ from $\varsigma$ through Meiosis with $c$ by (1)-(3).
        \STATE Obtain $me = \{me^t_{i}|i = 1, 2, ..., P;t \in \{A, B\}\}$ from $\tilde{\varsigma}$ through $f$ and $g$ by (4)-(8).
        \STATE Calculate loss $L$ by (9)-(12).
        \STATE Abate loss $L$ through optimizer updating parameters of $f$ and $g$.
    \UNTIL{all video clips are enumerated.}
\ENDFOR
\ENSURE base encoder $f$, throw away group projector $g$.
\end{algorithmic}
\end{algorithm}

\subsection{Fine-tuning Process}\label{HH}

To achieve excellent emotion classification performance, based on the learned feature representations, we further fine-tune the model using labeled samples. As shown in Fig. 1, we perform supervised training for emotion classification on a model composed of an initialized classifier and a basic encoder pre-trained with ME-NHACL.

We represent the training data as $ME$ and their labels as $y$. We represent the classifier as $c(\cdot)$. The label $y$ is a categorical variable. For example, if there are 4 emotion categories, $y$ can take 4 values: 0, 1, 2, or 3. We need to predict the emotion category $y$ for each sample $X \in R^{M \times C}$. The pre-trained basic encoder $f$ extracts representations from the raw ME signals $X$, which are used by the classifier $c(\cdot)$ to extract prediction features to obtain the predicted category $y^{\text{pre}} = c(f(X))$. We apply the cross-entropy function to define the loss function for the emotion classification task and apply an optimizer to minimize the loss function to optimize the model parameters. Finally, when the loss function converges, we obtain a model for predicting emotion recognition based on ME signals.

\section{Experimental} 
\label{section:Experimental}

In this section, we introduce the implementation details and experimental evaluation on the DEAP and HCI datasets. In the experiments, we compared ME-MHACL with other existing emotion recognition methods and evaluated its performance under limited labeled sample learning. By visualizing the feature representations learned by ME-MHACL, we explored the reasons for its effectiveness. By evaluating different combinations of hyperparameters, we explored meaningful patterns of the framework. In addition, we verified the rationality of the architecture design through control and ablation experiments.

\subsection{Implementation Detail}\label{AA}

In this section, we elaborate on the implementation details of the data set used in the experiment, data processing, and basic hyperparameters.

\subsubsection{Dataset}\label{A}

The widely used DEAP dataset~\cite{koelstra2011deap} includes ME signals recorded from 32 subjects while watching 40 one-minute music videos. The ME data of this dataset contains 32-channel EEG signals, 2-channel EOG signals, 2-channel EMG signals, 1-channel GSR signal, 1-channel respiration rate signal, 1-channel respiration belt pneumotachograph signal and 1-channel body temperature change signal, totaling 40 channels of valid data. Each trial data was recorded under rest for 3 seconds and stimulation for 60 seconds. The provider down-sampled the recorded 40-channel ME data to a sampling rate of 128hz and processed it with a band-pass filter in the frequency range of 4-45hz. After watching each video, the subjects were asked to rate each video on a scale of 1 to 9 for emotional arousal, valence, liking and dominance. We used arousal and valence scores for emotion recognition. We set the threshold values for arousal and valence ratings to 5. When the rating value is greater than 5.0, the corresponding ME signal is marked as high arousal or high valence. Otherwise, it is marked as low arousal or low valence. Each ME signal corresponds to two labels of valence and arousal, which can be used to construct 2 or 4 classification tasks. Table 2 shows the age and gender distribution of the samples in this dataset.

The MAHNOB-HCI dataset~\cite{soleymani2011multimodal} is an emotional dataset generated by 30 subjects watching 28 movie clips and 28 pictures. This dataset simultaneously collects relevant data with 6 cameras, a head-mounted microphone, an eye tracker and ME sensors. The ME data includes 32-channel EEG data, 3-channel ECG data, 1-channel GSR data on the finger, 1-channel skin temperature data (Temp), 1-channel respiration belt pneumotachograph signal (Resp) and 1 channel for marking the state. Each subject rated the Arousal and Valence values of each movie clip on a scale of 1-9. We converted the ratings into continuous values and used them as emotional labels. The ME signals contain valid data from 39 channels and down-sample the data to a sampling rate of 256Hz. We obtained labels in the two dimensions of arousal and valence by binarizing the evaluation values of arousal and valence, and these labels will be used in emotion recognition tasks based on the arousal dimension and valence dimension. Table 2 also shows the age and gender distribution of the samples in this dataset.

\begin{table}[!t]
\caption{Description of subjects and gender ratios based on age}
\begin{center}
\begin{tabular}{|c|c|c|c|c|}
\hline
\textbf{Dataset} & \textit{Subjects}& \textit{Age Distribution}& \textit{Male/Female}& \textit{Trials} \\
\hline
DEAP& 32& 22-37& 17/15& 40 \\
\hline
MAHNOB-HCI& 25& 19-40& 11/14& 20 \\
\hline
\end{tabular}
\label{tab：2}
\end{center}
\end{table}

\subsubsection{Data Process}\label{B}

On the DEAP dataset, we used a 1-second sliding window to separate the 63s signal of each trial into 63 non-overlapping ME signal segments. To improve accuracy, based on existing work ~\cite{koelstra2011deap}, we reduced the 3s resting state signal from the 60s emotional stimulation ME signal. In each trial, we averaged the 3s baseline ME signal segment to obtain a 1s average baseline ME signal segment~\cite{kan2022self}. The remaining 60 segments minus the average baseline segment become input samples. All samples correspond to a total of 2400 (40 60-second videos) repeated 1-second video clips. From the 2400 video clips, in a ratio of 70:15:15, 1680, 320 (actually should be 360) and 320 (actually should be 360) 1-second video clips were randomly divided into three groups. The three groups of video clips watched by the 32 subjects correspond to 53760, 11520 and 11520 (70:15:15) ME data segments, respectively, as training set, test set and validation set.

On the MAHNOB-HCI dataset, we first scaled and removed baseline drift for each channel of the ME signals~\cite{soleymani2011multimodal}. Similar to the DEAP dataset, we divided the movie videos into 1-second windows. Since the lengths of the test videos are different, we split adjacent windows along the time axis from front to back, removing the first three seconds of resting state signals from the 30-second emotional stimulation ME signals. In each trial, we averaged the 3s baseline ME signal segment to obtain a 1s average baseline ME signal segment. The remaining 24 segments minus the average baseline segment become input samples. From the 30 samples provided in the dataset, 5 samples without labels were removed. In a ratio of 70:15:15, each sample's 480 data segments with mean baseline removed were divided into 336, 72 and 72 (70:15:15) ME data segments, respectively as training set, test set and validation set.

\subsubsection{Basic Configuration}\label{C}

In order to accurately evaluate the performance of the pre-training framework for emotion recognition, we used two steps to evaluate the results. First, save the pre-trained model with different epochs. Next, select the model with the highest average accuracy of emotion recognition after 5 fine-tunings. This average accuracy is used as the result for evaluation.

In order to speed up the sampling, during the pre-training process, we set the five dimensions of the dataset tensor to correspond to video clips, subjects, 1, channels, and sliding window width, respectively. In the fine-tuning process, the first two axes of the dataset, video clips and subjects, are reshaped into a sample axis. The reshaped dataset's axes correspond to samples, 1, channels, and sampling points in turn. In the pre-training task, each epoch traverses each video clip of the dataset. A good pre-training task usually requires training for more than 2000 epochs. In order to reduce workload, we use the validation dataset to adjust the hyperparameters of the ME-MHACL framework and use the test dataset to evaluate the model. Listed in Table 3, $SHAPE_{tr}$, $SHAPE_{te}$, $SHAPE_{val}$ represent the tensor sizes of training test and validation datasets used for pre-training or fine-tuning. Epoch represents an appropriate number of pre-training or fine-tuning epochs to achieve good emotion recognition performance. Batchsize represents the number of samples in a small batch.

\begin{table*}[htbp]
\caption{HYPER PARAMETERS UTILIZED IN THE PROPOSED ME-MHACL}
\begin{center}
\begin{tabular}{|c|c|c|c|c|c|c|c|c|c|}
\hline
\textbf{$Dataset$} & \textit{$Link$}& \textit{$Epochs$}& \textit{$Batchsize$}& \textit{$LR$}& \textit{$\tau$}& \textit{$P$}& \textit{$Q$} & \textit{$SHAPE_{tr}$}& \textit{$SHAPE_{te/val}$}\\
\hline
\multirow{2}{*}{DEAP} & Pre-training& 4000& 32& $10^{-4}$& $10^{-1}$& 8& 2&$(1680,36,1,32,128)$& $(360,32,1,36,128)$ \\
\cline{2-10}
& Fine-tuning& 20& 2048& $10^{-3}$& -& -& -&$(53760,1,36,128)$& $(11520,1,36,128)$ \\
\hline
\multirow{2}{*}{MAHNOB-HCI} & Pre-training&475 & 32& $10^{-4}$& $10^{-1}$& 4& 8&$(336,25,1,36,256)$& $(72,25,1,36,256)$ \\
\cline{2-10}
& Fine-tuning& 20& 256& $10^{-3}$& -& -& -&$(8400,1,36,256)$& $(1800,1,36,256)$ \\
\hline
\end{tabular}
\label{tab:3}
\end{center}
\end{table*}

This paper implements experiments using PyTorch~\cite{NEURIPS2019_bdbca288} on an NVIDIA RTX2080ti GPU. The Adam optimizer~\cite{kingma2017adam} is used to minimize the loss function during the pre-training and fine-tuning processes. We denote the learning rate of the optimizer as lr. During the pre-training and fine-tuning processes, different values are applied for the number of iterations, batch size, temperature parameter $\tau$, learning rate lr, number of video clips per iteration $P$, number of samples per group $Q$, and tensor size of the dataset. As shown in Table II, we list all the hyperparameters used in the two processes on the DEAP and MAHNOB-HCI datasets.

\subsection{Emotion Classification Performance}\label{AA}
\subsubsection{Performance on DEAP}\label{A}

As shown in Table 4, on the DEAP dataset, due to the pioneering use of ME data, we can only compare with algorithms that use DEAP EEG data alone and algorithms that use EEG and EOG bimodal data. First, ME-MHACL is compared with two state-of-the-art methods on the two emotional dimensions of valence and arousal: MMResLSTM~\cite{ma2019emotion}, which uses multimodal data with residual long short-term memory networks, and ACRNN~\cite{tao2020eeg}, a hybrid network combining recurrent networks with channel attention mechanisms. From Table II, it can be seen that the proposed ME-MHACL is 2.67\% higher than the second highest in the valence dimension and 2.3\% higher than the second highest in the arousal dimension. The experimental results demonstrate the effectiveness of ME-MHACL in emotion recognition.

In order to verify the effectiveness of the proposed framework in the multimodal aspect, we first compared ME-MHACL with MindLink-Eumpy~\cite{li2021mindlink}, which is based on EEG and subject facial images collected from videos. ME-MHACL is 26.14\% higher than MindLink-Eumpy in the valence dimension and 34.44\% higher than MindLink-Eumpy in the arousal dimension. Secondly, we compared ME-MHACL with DCCA~\cite{liu2019multimodal}, which is based on EEG and EOG. ME-MHACL is 12.06\% higher than DCCA in the valence dimension and 7.8\% higher than DCCA in the arousal dimension. In addition to experiments with binary labels based on Valence and binary labels based on Arousal, we further compared them on a four-category classification problem: distinguishing four emotional labels: high valence and high arousal, high valence and low arousal, low valence and high arousal, low valence and high arousal, low valence and high arousal. In the four-category classification experiment, ME-MHACL is 0.84\% higher than DCCA. Once again, we compared ME-MHACL with GA-MLP~\cite{marjit2021eeg} based on EEG. ME-MHACL is 5.29\% higher than GA-MLP in the valence dimension, 2.42\% higher than GA-MLP in the arousal dimension, and 5.83\% higher than GA-MLP in the four-category classification experiment. In the dimensions of valence, arousal, and four categories, the average accuracy of ME-MHACL's fine-tuning scheme exceeds that of the self-supervised baseline scheme by 0.67\%, 1.33\%, and 0.06\%, respectively.

\begin{table}[!b]
\caption{Performance comparison on DEAP}
\begin{center}
\begin{tabular}{|c|c|c|c|}
\hline
\textbf{Method} & \textbf{\textit{Valence$^{\mathrm{a}}$}}& \textbf{\textit{Arousal$^{\mathrm{b}}$}}& \textbf{\textit{Four$^{\mathrm{c}}$}} \\
\hline
MindLink-Eumpy (2021)~\cite{li2021mindlink} & 70.25\% & 59.00\% & - \\
\hline
MMResLSTM (2019)~\cite{ma2019emotion} & 92.87\% & 92.30\% & - \\
\hline
ARCNN (2019)~\cite{tao2020eeg} & 93.72\% & 93.38\% & - \\
\hline
DCCA (2019)~\cite{liu2019multimodal} & 84.33\% & 85.62\% & 88.51\% \\
\hline
GA-MLP (2021)~\cite{marjit2021eeg} & 91.10\% & 91.02\% & 83.52\% \\
\hline
ME-MHACL (Fully-supervised) & 95.72\% & 92.11\% & 89.29\% \\
\hline
ME-MHACL (Fine-tuning) & 96.39\% & 93.44\% & 89.35\% \\
\hline
\multicolumn{4}{l}{%
    \shortstack[l]{$^{\mathrm{a}}$Average accuracy over five trials in Binary classification based on\\Valence.}%
}\cr
\multicolumn{4}{l}{%
    \shortstack[l]{$^{\mathrm{b}}$Average accuracy over five trials in Binary classification based on\\Arousal.}%
}\cr
\multicolumn{4}{l}{%
    \shortstack[l]{$^{\mathrm{c}}$Average accuracy over five trials in four categories based on\\Valence and Arousal.}%
}\cr
\end{tabular}
\label{tab:4}
\end{center}
\end{table}

Comparing the above data, we can find that whether it is the experiment of a single dimension of Arousal or Valence, or the four-category classification experiment, ME-MHACL has better performance than DCCA, proving that ME-MHACL can more effectively extract emotional features that adapt to sample differences in ME data compared to other modal information.

\begin{figure*}[htbp]
	\centering
	\subfigure[]{\includegraphics[scale=0.5]{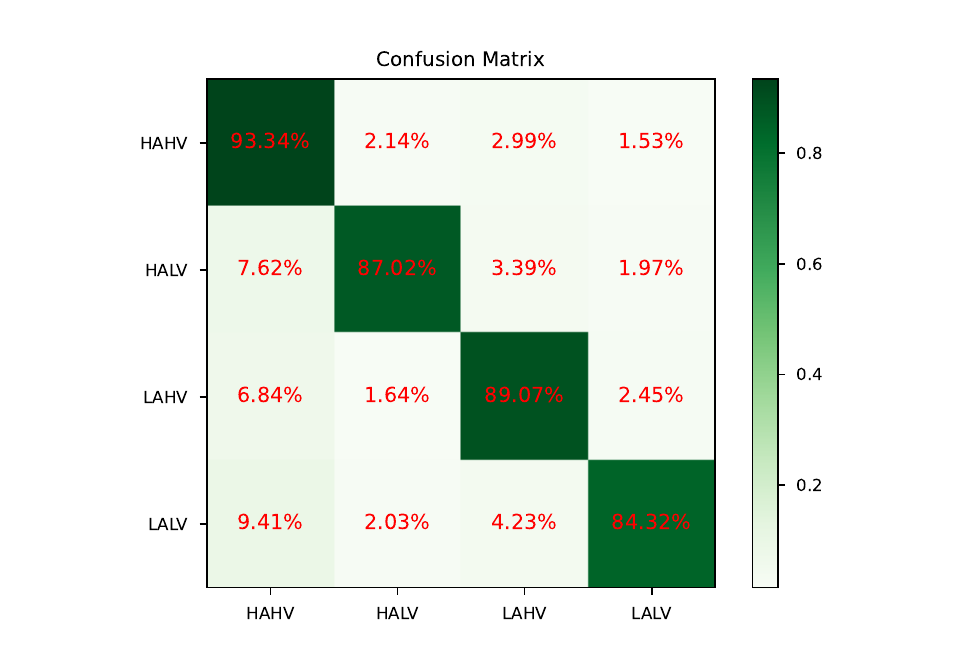}}
	\hspace{10 pt}
	\subfigure[]{\includegraphics[scale=0.5]{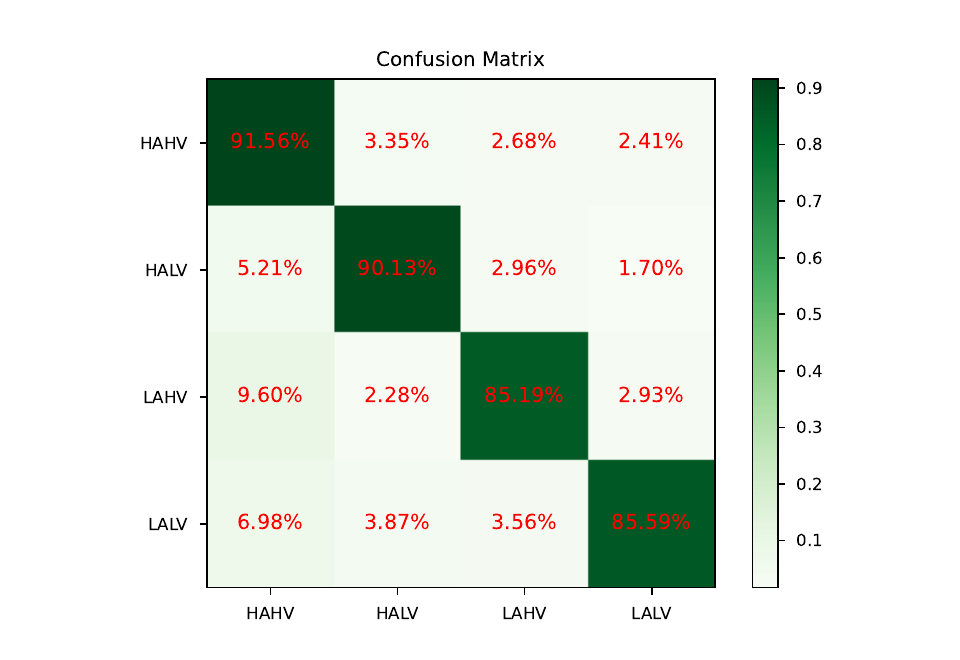}}
	\caption{Four categories of performance comparison using DEAP. (a) is the ME-MHACL fine-tuning scheme quad-classification confusion matrix; (b) is a four-category confusion matrix for the self-supervised baseline scenario.}
	\label{fig:6}
\end{figure*}

Comparing the confusion matrices of the four-category classification experiments of DEAP shown in Fig. 6 for ME-MHACL's fine-tuning structure and self-supervised baseline, ME-MHACL's fine-tuning structure performs better in the case of high arousal and high valence and in the case of low arousal and high valence, while ME-MHACL's self-supervised scheme performs better in the case of high arousal and high valence and in the case of high arousal and low valence. In the case of changes in data channels, ME-MHACL's fine-tuning scheme can still maintain stable inference performance, proving that ME-MHACL has generalization in cross-dataset experiments.

\subsubsection{Performance on MAHNOB-HCI}\label{B}

As shown in Table 5, on the MAHNOB-HCI dataset, our proposed ME-MHACL is first compared with TSception~\cite{ding2022tsception}, which is based on single-modal EEG data: ME-MHACL is 33.62\% higher than TSception in the valence dimension and 35.28\% higher than DCCA in the arousal dimension. Then, ME-MHACL is compared with MINDLINK\_EUMPY~\cite{li2021mindlink}, which is based on EEG and facial video screenshots: in the valence dimension, ME-MHACL is 16.33\% higher than MINDLINK-EUMPY; in the arousal dimension, ME-MHACL is 18.89\% higher than MINDLINK-EUMPY. Finally, ME-MHACL is compared with HCNNS-MFB~\cite{zhang2022emotion}, which is also based on single-modal EEG data: ME-MHACL is 4.72\% higher than HCNNS-MFB in the valence dimension and 5.94\% higher than HCNNS-MFB in the arousal dimension. By observing Table IV, we can see that ME-MHACL's model fine-tuning scheme and self-supervised learning algorithm have a significant advantage over other multimodal algorithms and single-modal algorithms, and ME-MHACL can also achieve an average accuracy of over 93.5\% in the four-category classification experiment. In the dimensions of valence, arousal, and four categories, the average accuracy of ME-MHACL's fine-tuning scheme exceeds that of the self-supervised baseline scheme by 0.06\%, 0.22\%, and 0.11\%, respectively.

\begin{table}[!b]
\caption{Performance comparison on MAHNOB-HCI}
\begin{center}
\begin{tabular}{|c|c|c|c|}
\hline
\textbf{Method} & \textbf{\textit{Valence$^{\mathrm{a}}$}}& \textbf{\textit{Arousal$^{\mathrm{b}}$}}& \textbf{\textit{Four$^{\mathrm{c}}$}} \\
\hline
TSception (2021)~\cite{ding2022tsception} & 61.27\% & 60.61\% & - \\
\hline
MindLink-Eumpy (2021)~\cite{li2021mindlink} & 78.56\% & 77.22\% & - \\
\hline
HCNNS-MFB (2022)~\cite{zhang2022emotion} & 90.17\% & 90.17\% & - \\
\hline
ME-MHACL (Fully-supervised) & 94.83\% & 95.89\% & 93.61\% \\
\hline
ME-MHACL (Fine-tuning) & 94.89\% & 96.11\% & 93.72\% \\
\hline
\multicolumn{4}{l}{%
    \shortstack[l]{$^{\mathrm{a}}$Average accuracy over five trials in Binary classification based on\\Valence.}%
}\cr
\multicolumn{4}{l}{%
    \shortstack[l]{$^{\mathrm{b}}$Average accuracy over five trials in Binary classification based on\\Arousal.}%
}\cr
\multicolumn{4}{l}{%
    \shortstack[l]{$^{\mathrm{c}}$Average accuracy over five trials in four categories based on\\Valence and Arousal.}%
}\cr
\end{tabular}
\label{tab:4}
\end{center}
\end{table}

Thus, it can be seen that ME-MHACL can obtain effective representations based on the valence and arousal dimensions in the multimodal data of MAHNOB-HCI. The four-category classification experiment further proves the robustness of ME-MHACL in the challenging task of MAHNOB-HCI.

In this experiment, the data modalities used in MAHNOB-HCI and DEAP are the same: EEG, GSR, Resp, and Temp. However, the data of MAHNOB-HCI has one less channel of GSR data than that of DEAP. Based on the above cross-dataset differences, we conducted a four-category classification experiment based on the dual labels of Valence and Arousal.

\begin{figure*}[!b]
	\centering
	\subfigure[]{\includegraphics[scale=0.5]{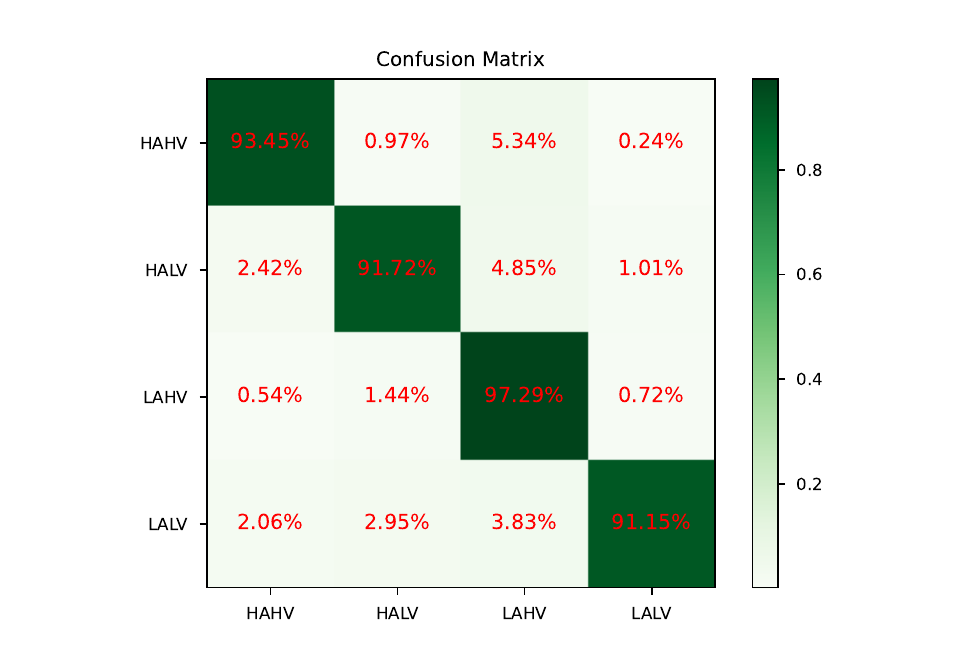}}
	\hspace{10 pt}
	\subfigure[]{\includegraphics[scale=0.5]{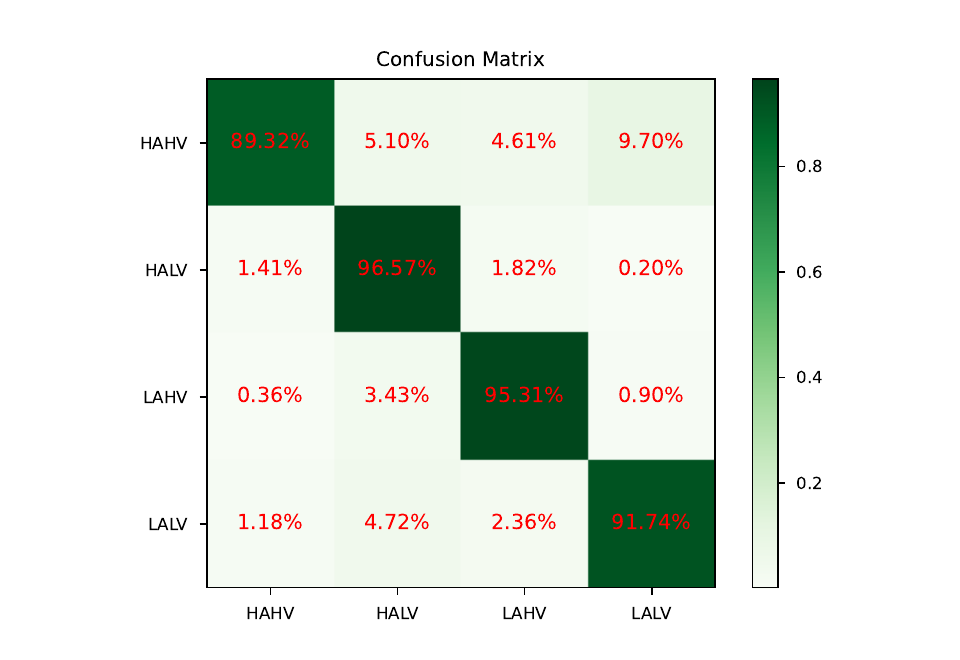}}
	\caption{Four-class performance comparison using MAHNOB-HCI. (a) is the ME-MHACL fine-tuning scheme quad-classification confusion matrix; (b) is a four-category confusion matrix for the self-supervised baseline scenario.}
	\label{fig:7}
\end{figure*}

Comparing the confusion matrices of the four-category classification experiments of MAHNOB-HCI shown in Fig. 7 for ME-MHACL's fine-tuning structure and self-supervised baseline, ME-MHACL's fine-tuning structure performs better in the case of high arousal and high valence and in the case of low arousal and low valence, while ME-MHACL's self-supervised scheme performs better in the case of high arousal and low valence and in the case of low arousal and low valence. In the case of changes in data channels, ME-MHACL's fine-tuning scheme can still maintain stable inference performance, proving that ME-MHACL has generalization in cross-dataset experiments.

\subsection{The experiments based on data modes}\label{CC}

In this section, we conduct ablation experiments by selecting data modalities. Firstly, we investigate the impact of the number of channels in ME data on the accuracy of emotion recognition. Secondly, we study the effect of the combination of ME data on the accuracy of emotion recognition. Finally, we use attention heatmaps to visualize the attention of each channel data for each modality. 

\subsubsection{Data mode ablation experiment}\label{A}

\begin{figure*}[!t]
	\centering
	\subfigure[]{\includegraphics[scale=0.5]{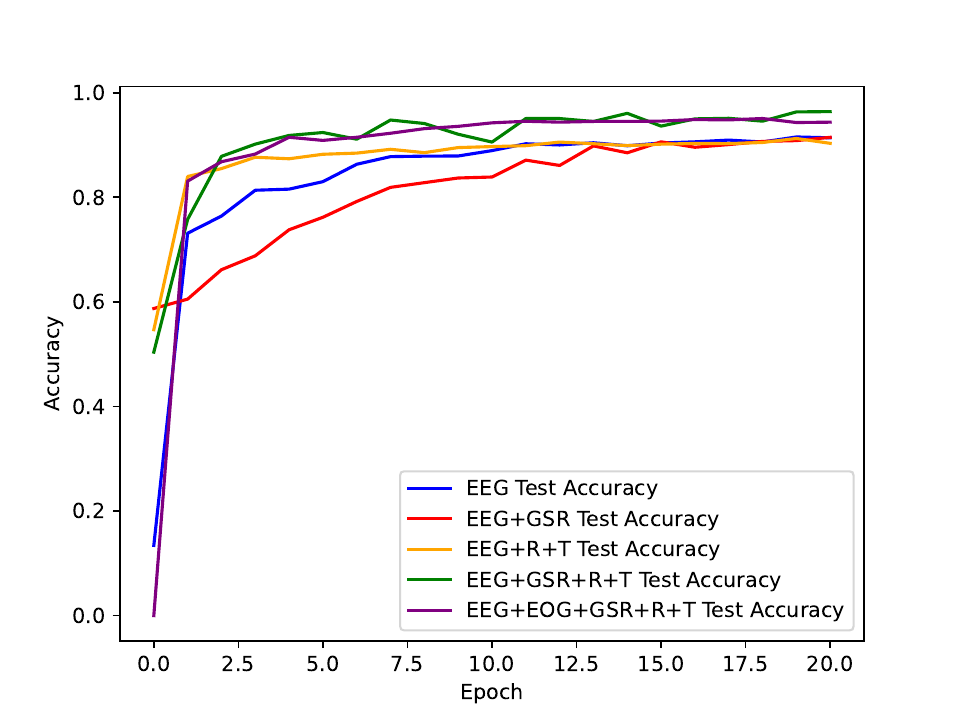}}
	\hspace{10 pt}
	\subfigure[]{\includegraphics[scale=0.5]{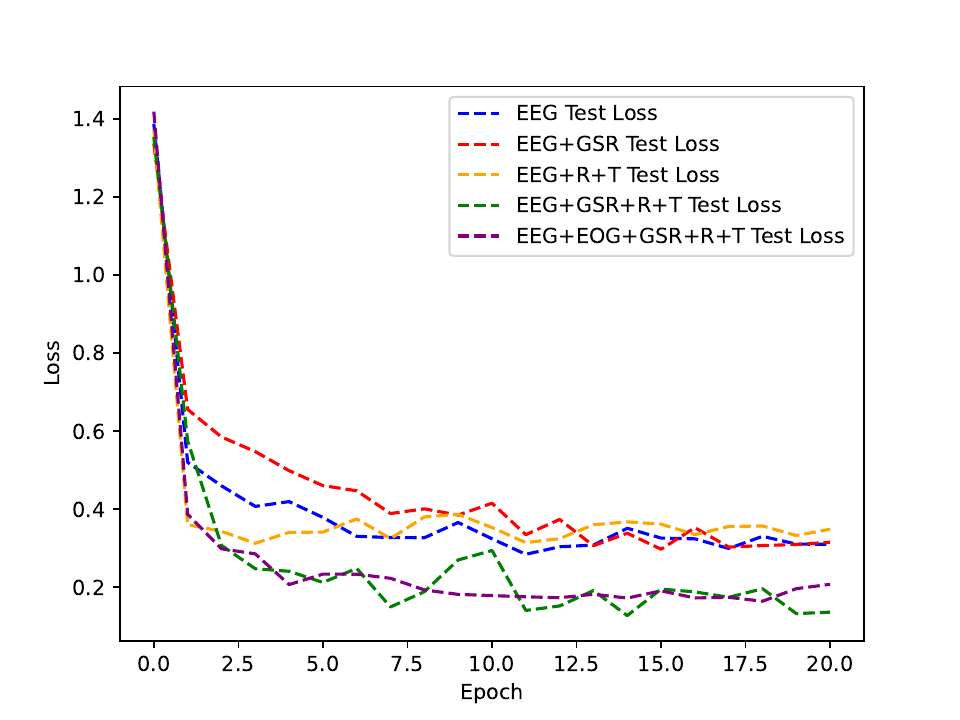}}
	\caption{Four-class performance comparison using MAHNOB-HCI. (a) is the ME-MHACL fine-tuning scheme quad-classification confusion matrix; (b) is a four-category confusion matrix for the self-supervised baseline scenario.}
	\label{fig:8}
\end{figure*}

In this experiment, we fine-tuned the binary classification of Valence based on the DEAP dataset by adjusting the number of channels of ME data. In each experiment of this group, EEG signals were used as the main modality. As shown in Fig.~\ref{fig:8}, fine-tuning with 32-channel EEG unimodal data resulted in an average test accuracy of 91.38\% and an average loss of 31.01\%. Fine-tuning with 32-channel EEG and 1-channel GSR bimodal data resulted in an average test accuracy of 91.47\% and an average loss of 31.57\%. Fine-tuning with 32-channel EEG, 1-channel Resp, and 1-channel Temp trimodal data resulted in an average test accuracy of 90.32\% and an average loss of 34.88\%. Fine-tuning with 32-channel EEG, 1-channel GSR, 1-channel Resp, and 1-channel Temp quadmodal data resulted in an average test accuracy of 96.39\% and an average loss of 13.64\%. Fine-tuning with 32-channel EEG, 2-channel EOG, 1-channel GSR, 1-channel Resp, and 1-channel Temp pentamodal data resulted in an average test accuracy of 94.37\% and an average loss of 20.77\%, and the above results were obtained by averaging over five experiments.

Although the overall trend shows that as the number of data modalities increases, the accuracy of emotion recognition remains on an overall upward trend. However, due to the domain specificity of specific tasks, the effectiveness of data is implicitly related to the task, and it is not positively correlated with the number of channels of ME data.

\subsubsection{Dual mode combination comparison}\label{B}

\begin{figure}[!b]
\centerline{{\includegraphics[scale=0.5]{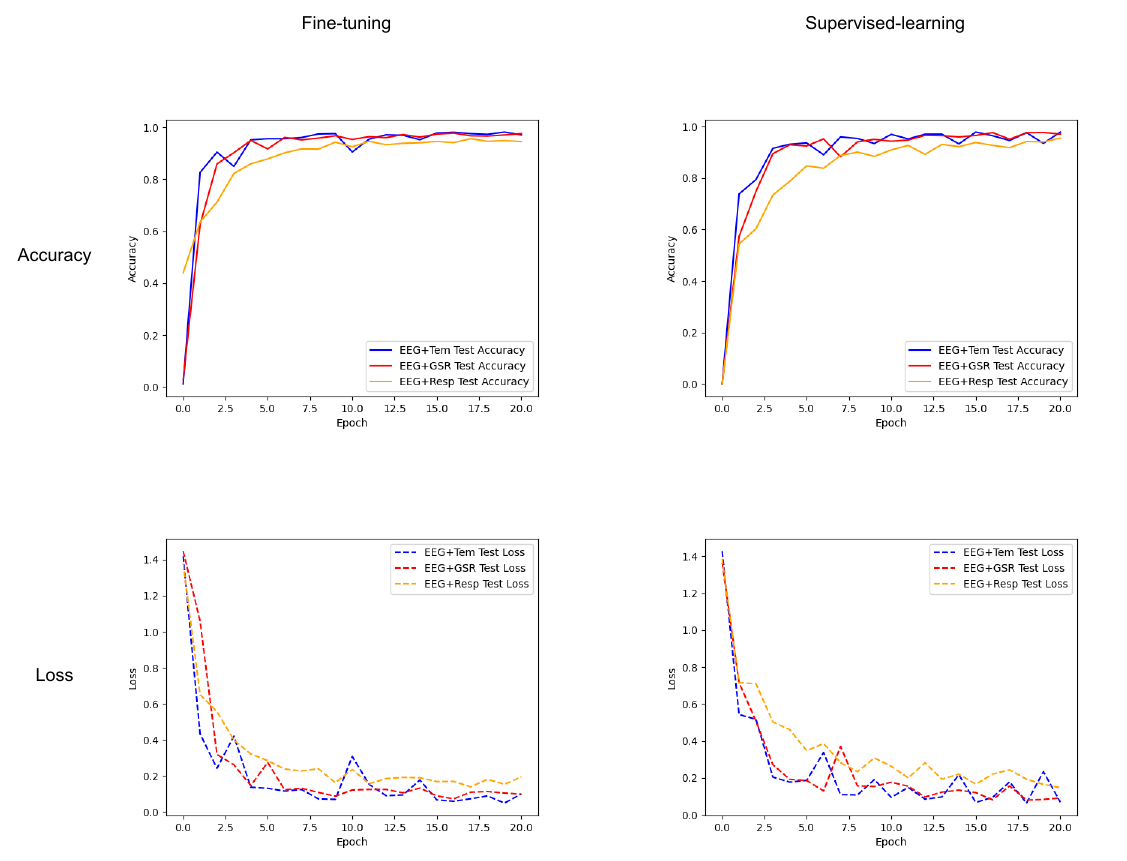}}}
\caption{In the fine-tuning and self-supervised comparison experiment based on the valence label, in the binary classification task based on the valence label, whether it is fine-tuning or self-supervised, the effect of using the EEG and temperature combination is basically equal to the effect of using the EEG and skin resistance combination. However, the EEG and respiration rate combination has a lower recognition rate based on the valence label and a higher loss rate.}
\label{fig:9}
\end{figure}

In this experiment, we tested the binary classification of Valence based on different combinations of ME data from the MAHNOB\_HCI dataset, including combinations of EEG and temperature, EEG and skin resistance, and EEG and respiration rate.As shown in Fig.~\ref{fig:9}, we first compared the accuracy of emotion recognition for fine-tuning and self-supervised learning of three bimodal data combinations in the first row. In the fine-tuning stage, the accuracy of the EEG and temperature combination was 97.06\%; the accuracy of the EEG and skin resistance combination was 97.56\%; and the accuracy of the EEG and respiration rate combination was 94.5\%. In self-supervised learning, the accuracy of the EEG and temperature combination was 97.89\%; the accuracy of the EEG and skin resistance combination was 97.11\%; and the accuracy of the EEG and respiration rate combination was 95.56\%. Then, in the second row, we compared the loss rate of emotion recognition for fine-tuning and self-supervised learning of three bimodal data combinations. In the fine-tuning stage, the loss rate of the EEG and temperature combination was 10.02\%; the loss rate of the EEG and skin resistance combination was 9.93\%; and the loss rate of the EEG and respiration rate combination was 19.48\%. In self-supervised learning, the loss rate of the EEG and temperature combination was 6.87\%; the loss rate of the EEG and skin resistance combination was 9.16\%; and the loss rate of the EEG and respiration rate combination was 14.73\%.

In the binary classification experiment based on arousal using bimodal data, it also includes combinations of EEG and temperature, EEG and skin resistance, and EEG and respiration rate. As shown in Fig. 10, firstly, we compared the accuracy of emotion recognition for fine-tuning and self-supervised learning of three bimodal data combinations in the first row. In the fine-tuning stage, the accuracy of the EEG and temperature combination was 96.39\%; the accuracy of the EEG and skin resistance combination was 97.62\%; and the accuracy of the EEG and respiration rate combination was 94.39\%. In self-supervised learning, the accuracy of the EEG and temperature combination was 97.33\%; the accuracy of the EEG and skin resistance combination was 96.94\%; and the accuracy of the EEG and respiration rate combination was 95.56\%. Then, in the second row, we compared the loss rate of emotion recognition for fine-tuning and self-supervised learning of three bimodal data combinations. In the fine-tuning stage, the loss rate of the EEG and temperature combination was 13.48\%; the loss rate of the EEG and skin resistance combination was 9.23\%; and the loss rate of the EEG and respiration rate combination was 19.71\%. In self-supervised learning, the loss rate of the EEG and temperature combination was 6.87\%; the loss rate of the EEG and skin resistance combination was 9.48\%; and the loss rate of the EEG and respiration rate combination was 14.7\%.

\begin{figure}[!t]
\centerline{{\includegraphics[scale=0.5]{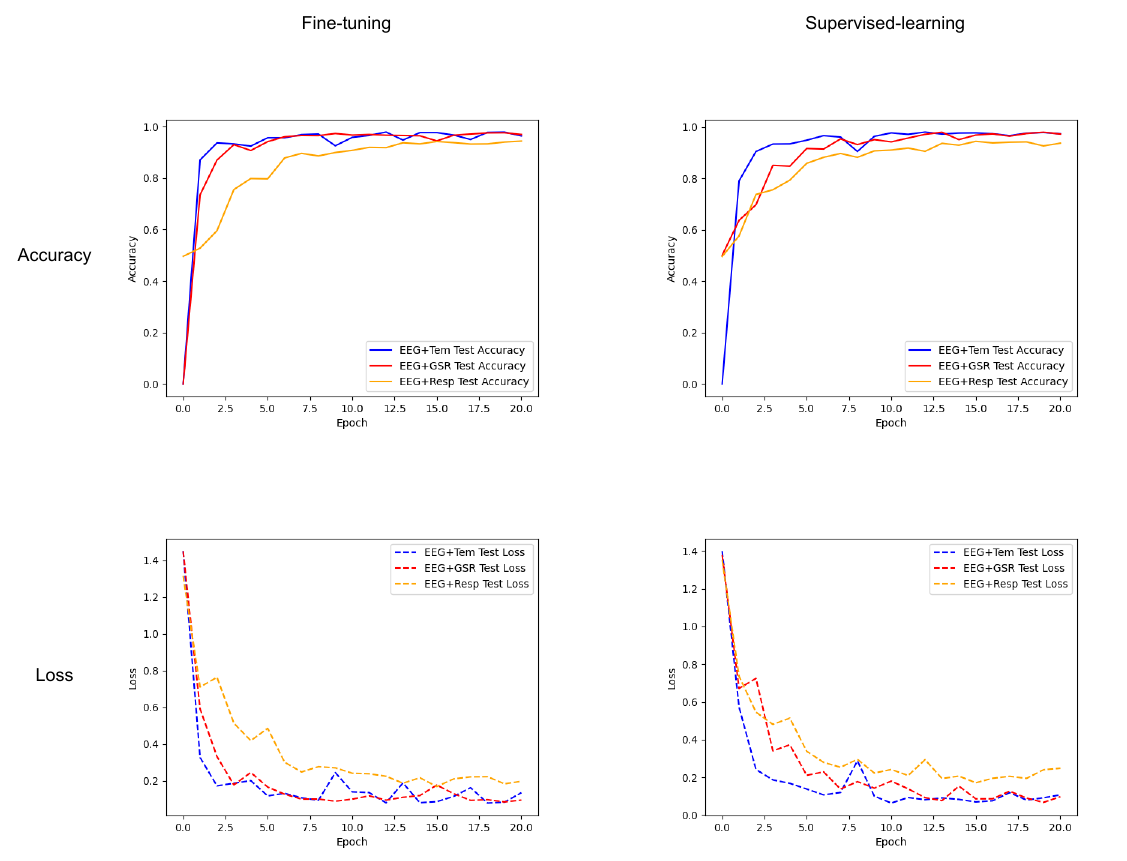}}}
\caption{In the fine-tuning and self-supervised comparison experiment based on the valence label, in the binary classification task based on the valence label, whether it is fine-tuning or self-supervised, the effect of using the EEG and temperature combination is basically equal to the effect of using the EEG and skin resistance combination. However, the EEG and respiration rate combination has a lower recognition rate based on the valence label and a higher loss rate.}
\label{fig:10}
\end{figure}

Through the above experiments, we found that in emotion recognition tasks based on different labels, different combinations of bimodal data have different effects on the emotion recognition of Arousal and valence. Lang et al.~\cite{lang1993looking} found that the average value of GSR is related to the level of arousal, and slow breathing is related to relaxation, while irregular rhythm, rapid breathing, and cessation of breathing are related to stronger emotions such as anger or fear. [DEAP] When the subject is in a state of anger or fear, their skin temperature may rise and their breathing rate may increase; when the subject's emotions are relaxed, their breathing may slow down~\cite{koelstra2011deap}. The above phenomenon is the origin of the difference in effective representation between different modal electrophysiological data and emotion recognition tasks.

\subsection{Ablation experiments based on MHA}\label{CC}

In this section, we will compare the representation performance of tensors using multi-head attention mechanism in group projectors in binary classification and four-class classification tasks.

\subsubsection{Four-classification T-SNE visualization based on MHA}\label{A}

As shown in Fig. 11, based on the MAHNOB-HCI dataset, according to the self-supervised four-class task mentioned earlier, the data of four modalities is mapped to a two-dimensional plane through the 1800-dimensional features extracted by the multimodal basic encoder using t-SNE~\cite{van2008visualizing}. After 10 samples of self-supervised learning, we selected three sampling results to obtain the three images in the first row. After the 1800-dimensional feature representation processed by the multimodal multi-head attention mechanism, we also sampled three times to obtain the three images in the second row.

\begin{figure*}[htbp]
\centerline{{\includegraphics[scale=0.6]{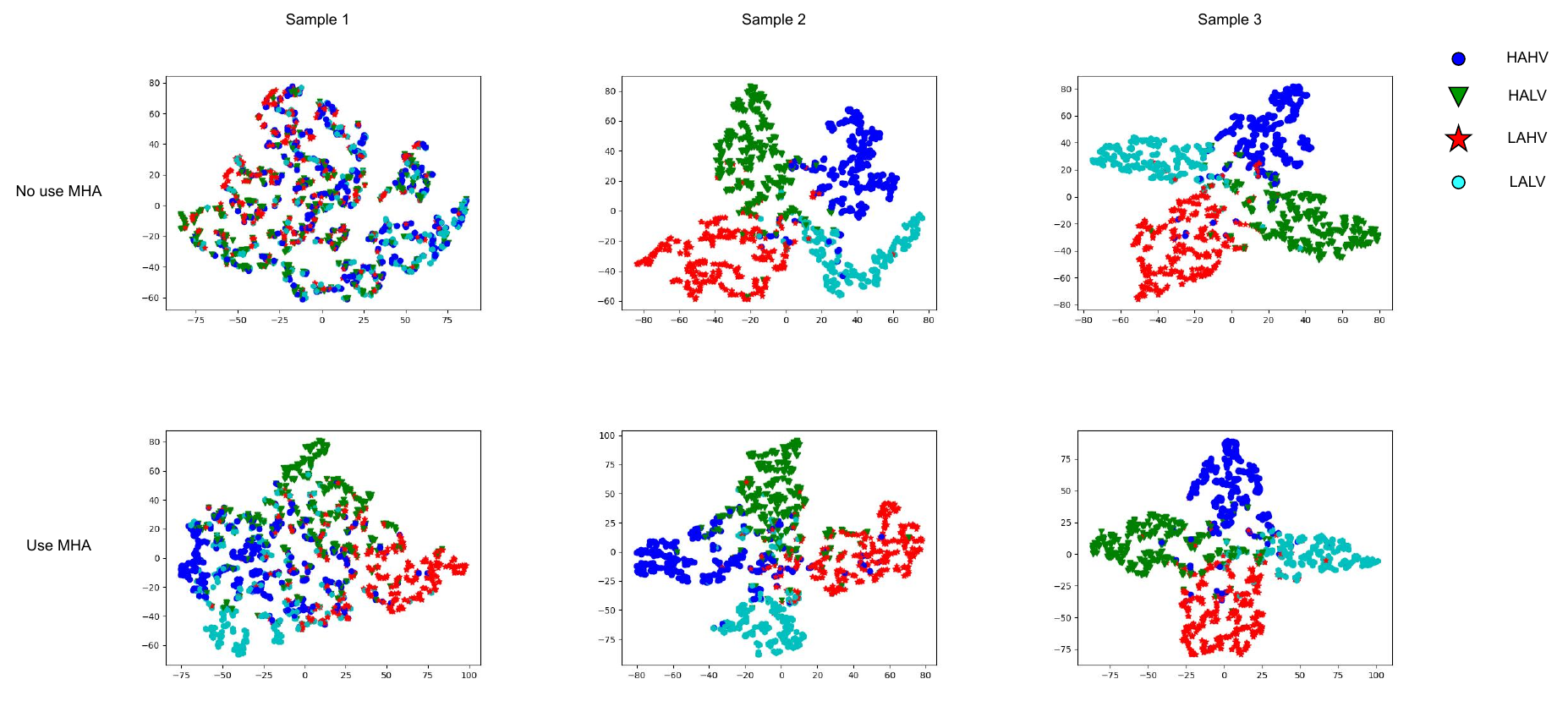}}}
\caption{In the fine-tuning and self-supervised comparison experiment based on the valence label, in the binary classification task based on the valence label, whether it is fine-tuning or self-supervised, the effect of using the EEG and temperature combination is basically equal to the effect of using the EEG and skin resistance combination. However, the EEG and respiration rate combination has a lower recognition rate based on the valence label and a higher loss rate.}
\label{fig:11}
\end{figure*}

\subsubsection{Two-classification T-SNE visualization based on MHA}\label{B}

As shown in Fig. 12, based on the MAHNOB-HCI dataset, according to the binary classification task of the Valence label, the data of four modalities is mapped to a two-dimensional plane through the 1800-dimensional features extracted by the multimodal basic encoder using t-SNE. After 10 samples of self-supervised learning, we selected three sampling results to obtain the three images in the first row. After the 1800-dimensional feature representation processed by the multimodal multi-head attention mechanism, we also sampled three times to obtain the three images in the second row.

\begin{figure*}[!b]
\centerline{{\includegraphics[scale=0.65]{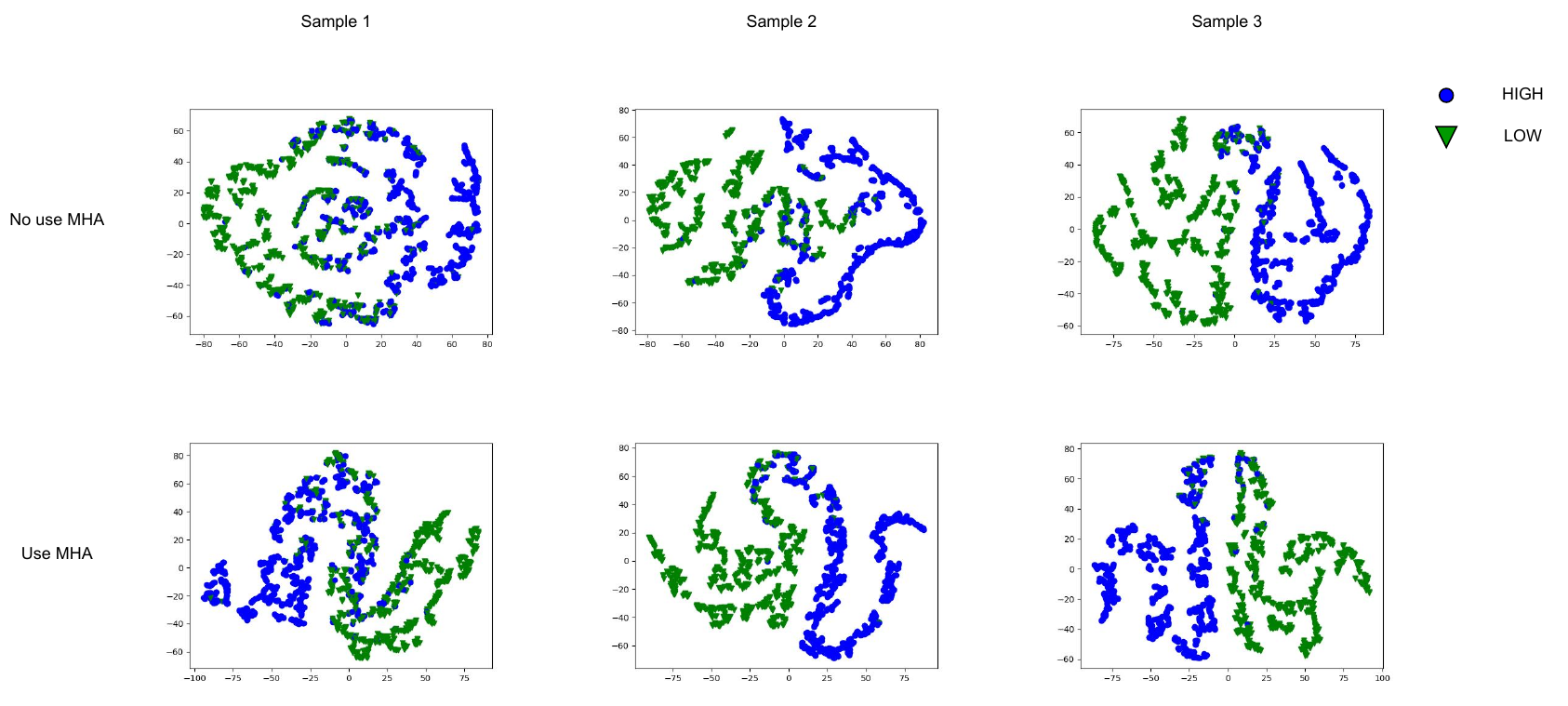}}}
\caption{In the fine-tuning and self-supervised comparison experiment based on the valence label, in the binary classification task based on the valence label, whether it is fine-tuning or self-supervised, the effect of using the EEG and temperature combination is basically equal to the effect of using the EEG and skin resistance combination. However, the EEG and respiration rate combination has a lower recognition rate based on the valence label and a higher loss rate.}
\label{fig:12}
\end{figure*}

Whether it is a binary classification task based on Arousal or Valence, or a four-class classification task based on two-dimensional labels, the multi-head attention module in the multimodal group projector of ME-MHACL not only learns stimulus-related feature representations, but also enables the model to distinguish whether different stimuli come from continuous videos. Without using the multi-head attention module, there are more indistinguishable representations where different emotion labels are mixed together. When using the multi-head attention module, there are fewer feature representations where different emotion labels are mixed together, showing better emotion discrimination ability. This reflects that the multi-head attention module in the multimodal group projector enables ME-MHACL to learn video-level stimulus-related representations, thereby improving emotion recognition performance. 

\section{Conclusion and prospect} 
\label{section:Conclusion and prospect}

This paper proposes a multimodal emotion recognition method based on multi-head attention mechanism, which can effectively use ME signals for emotion recognition and fully mine the complementary information between electroencephalogram (EEG), skin resistance (GSR), respiration rate (Respiration), and temperature (Temperature). Experiments were conducted on two publicly available multimodal emotion datasets, and the results show that the proposed method outperforms existing benchmark methods in terms of accuracy and stability of emotion prediction, and can better distinguish different emotional states. 

There are several aspects of this work that can be further improved and expanded:
\begin{itemize}
\item This paper only considers four modalities of ME data, and in the future, other modalities of data, such as facial expressions, voice, body posture, etc., can be introduced to enhance the effectiveness and robustness of emotion recognition.
\item This paper only uses two emotional dimensions, arousal and valence, and in the future, more emotional dimensions, such as dominance, value, expectation, etc., can be considered to more comprehensively describe emotional states.
\item This paper only uses static emotion labels, and in the future, dynamic emotion labels, such as continuous emotion curves and changing emotion intensity, can be considered to more realistically reflect the process of emotion change.
\end{itemize}

\bibliographystyle{IEEEtran}
\bibliography{refs}

\end{document}